\documentclass[11pt]{article}
\usepackage{hyperref}
\pdfoutput=1
\begin{document}
\title{Sand swimming lizard: sandfish}
\author{Ryan D. Maladen$^1$, Yang Ding$^2$, Adam Kamor$^2$, Daniel I. Goldman$^{1,2}$
\\\vspace{6pt} $^1$ Bioengineering Program, $^2$ School of Physics, Georgia Institute of Technology, USA.
}

\maketitle
\begin{abstract}
In this fluid dynamics video, we use high-speed x-ray imaging to reveal how a small ($\sim$ 10cm) desert dwelling lizard, the sandfish ({\em Scincus scincus}), swims within a granular medium \footnote{
 Maladen, R.D., Ding, Y., Li, C., and Goldman, D.I., Undulatory Swimming in Sand: Subsurface Locomotion of the Sandfish Lizard, \textbf{Science}, 325, 314, 2009}. On the surface, the lizard uses a standard diagonal gait, but once below the surface, the organism no longer uses limbs for propulsion. Instead it propagates a large amplitude single period sinusoidal traveling wave down its body and tail to propel itself at speeds up to $\approx 1.5$ body-length/sec. Motivated by these experiments we study a numerical model of the sandfish as it swims within a validated soft sphere Molecular Dynamics granular media simulation. We use this model as a tool to understand dynamics like flow fields and forces generated as the animal swims within the granular media.
\end{abstract}

The link to the video is:
\href{http://ecommons.library.cornell.edu/bitstream/1813/14104/2/Maladen_Sandfish_mpg1.mpg}{Video1-mpg1 format} 

\href{http://ecommons.library.cornell.edu/bitstream/1813/14104/3/Maladen_Sandfish_mpg2.mpg}{Video2-mpg2 format} 
\end{document}